\documentclass[]{aa}

\usepackage{graphicx}
\usepackage{amsmath}

\usepackage{txfonts}

\begin{document}

\title{An alternative look at the snowline in protoplanetary 
disks}

\author{K. Kornet\inst{1} \and M. R{\'o}{\.z}yczka\inst{1} \and  T.F. 
Stepinski\inst{2}}

\offprints{K. Kornet, \email{kornet@camk.edu.pl}}

\institute{Nicolaus Copernicus Astronomical Center , 
           Bartycka 18  , Warsaw, PL-00-716, Poland\\ 
           \email{kornet@camk.edu.pl,mnr@camk.edu.pl}
\and
Lunar and Planetary Institute, 3600 Bay Area Blvd., Houston, TX 
77058, USA\\\email{tom@lpi.usra.edu}
}

\date{Received 2 July 2003 / Accepted 12 November 2003}

\titlerunning{An alternative look at the snowline}
\authorrunning{Kornet, R{\'o}{\.z}yczka \& Stepinski}

\msnr{AA/2003/0036}

\abstract{ We have calculated an evolution of protoplanetary disk from
  an extensive set of initial conditions using a time-dependent model
  capable of simultaneously keeping track of the global evolution of
  gas and water-ice.  A number of simplifications and idealizations
  allows for an embodiment of gas-particle coupling, coagulation,
  sedimentation, and evaporation/condensation processes. We have shown
  that, when the evolution of ice is explicitly included, the location
  of the snowline has to be calculated directly as the inner edge of
  the region where ice is present and not as the radius where disk's
  temperature equals the evaporation temperature of water-ice. The
  final location of the snowline is set by an interplay between all
  involved processes and is farther from the star than implied by the
  location of the evaporation temperature radius. The evolution
  process naturally leads to an order of magnitude enhancement
  in surface density of icy material.   }

\maketitle

\section{Introduction }

Water-ice in the protoplanetary disk can exist beyond a minimum
distance from the star called the snowline. Although the snowline is
defined directly by the presence of ice, the standard method of
calculating its location is indirect and uses disks temperatures.
Thus, in the present literature, the snowline is the location where
the disks temperature is equal to the sublimation/condensation
temperature of water-ice.  The total surface density of all solids,
$\Sigma_s$, increases rapidly outside the snowline because water-ice,
the most abundant species of solid, becomes available and its
contribution dominates the value of $\Sigma_s$. In the core accretion
- gas capture scenario of giant planet formation high values of
$\Sigma_s$ are necessary to produce solid cores on time scales
consistent with the presence of a gaseous nebula. Thus, at least in
such a scenario, the importance of the snowline derives from a notion
that it marks the inner edge of the giant planet formation zone.

In the context of the Solar System formation, the so-called
``minimum-mass solar nebula'' model has been used frequently for
quantitative analysis. In this steady-state model the temperature is
derived from radiative equilibrium with the solar radiation field
(Hayashi \cite{Hayashi}) and the snowline is located at a distance
$R_{sl}=2.7$ AU from the Sun. Just the presence of ice exterior to
$R_{sl}$ increases solid abundance by a factor of 4 yielding
$\Sigma_s=6.8$ g cm$^{-2}$ at the snowline and $\Sigma_s=2.7$ g
cm$^{-2}$ at $r=5$ AU (present, and, by assumption, the original
location of Jupiter). However, even with addition of ice, the values
of $\Sigma_s$ are still too low for a rapid formation of giant
planets cores. Lissauer (\cite{Lissauer}) calculated that surface
density of solids 5 -- 10 times greater than this given by the Hayashi
model are required to grow the Jovian core fast enough to accrete the
gaseous envelope before the solar nebula is dispersed.  Stevenson \&
Lunine (\cite{Stevenson}) proposed a mechanism for a further
enhancement of the abundance of water-ice just exterior to the
snowline by diffusive redistribution of water vapor through the
snowline. They calculated that such a mechanism can enhance the value
of $\Sigma_s$ by 1 to 2 orders of magnitude, making the rapid
formation of a Jovian core possible at the location of the snowline.
However, diffusive redistribution cannot explain core formation for
other giant planets in the Solar System.

The discovery of $\sim 100$ giant extrasolar planets (for a recent
review see Bodenheimer \& Lin \cite{Bodenheimer02}) posed formidable
theoretical problems for the traditional scenario of giant planet
formation because most exoplanets were found interior to the snowline
as calculated from the ``minimum mass'' model, and a significant
fraction of those (the so-called {\it hot Jupiters}) - closer than 0.1
AU, where both high temperature and insufficient amount of matter
should have prevented their formation. It appears that post-formation
orbital evolution such as migration of a planet under the influence of
tidal torque from the disk matter (see e.g. Lin et al. \cite{Lin00})
or gravitational close encounters between planets leading to
significant changes of their orbits (see e.g. Rasio \& Ford
\cite{Rasio96}) need to be invoked to explain the orbital
properties of hot Jupiters. This notwithstanding, a revisiting of the
snowline concept is also needed to understand the original formation of
giant exoplanets, and to explain orbital properties of those
exoplanets that seem to escape a significant post-formation orbital
evolution.

Two important developments call for the re-evaluation of the snowline
concept.  First, it becomes clear from astronomical observations (for
a compact review see Beckwith \cite{Beckwith}) that protoplanetary
disks, of which a solar nebula is thought to be a particular example,
are active, ever-changing entities with limited life-spans that cannot
be modeled successfully by a steady-state, phenomenological model like
the ``minimum mass'' model. Instead, numerical and analytical models
based on the concept of time-dependent accretion disk theory have been
developed (Ruden \& Pollack \cite{Ruden}, Sterzik \& Morfill
\cite{Sterzik}, Stepinski \cite{Stepinski}). In such models different
initial conditions lead to disks of different character and
evolutionary history, offering the potential to explain observed
diversity in exoplanets properties. In an evolving disk, the location
of the sublimation/condensation radius changes, and it has to be
considered as a function of time $R_{evap}(t)$. This function is
further parameterized by initial conditions.

Second, it has been recognized that the solid particles evolve
differently to the gas (see e.g.  Weidenschilling \& Cuzzi
\cite{Weidenschilling93} and references therein). The differences
between global evolution of gaseous and solid components of the
protoplanetary disk have been calculated by Stepinski \& Valageas
(\cite{SV96}, \cite{SV97}). They have shown that in the evolving
protoplanetary disk, the sub-disk of water-ice solids decouples over
time from the gaseous disk leading to significant departures from a
constant solids-to-gas mass ratio The ice sub-disk evolves towards a
swarm of icy planetesimals characterized by the value of $\Sigma_s$
considerably higher than at the beginning of the evolution. Within
such a model the standard method of calculating the location of the
snowline is not useful. Because the gas and the solids decouple, the
snowline location cannot be based on the temperature of the gas.
Instead, the snowline $R_{sl}(t)$ should be calculated directly from
its original definition as the inner edge of the region where
water-ice is present.

Recently, the location of the snowline has been recalculated.
Sasselov \& Lecar (\cite{Sasselov}) considered passive and low
accretion rate models of a protoplanetary disk instead of the
``minimum mass'' model to calculate the location of the snowline using
a standard technique of equating $R_{sl}$ with $R_{evap}$. They
obtained $R_{sl}\approx 1$ AU, significantly closer to a star than a
``distant'' snowline at 2.7 AU as predicted by the ``minimum mass''
model. The focus of the Sasselov \& Lecar work was on using a model of the
disk that is more consistent with observations than the ``minimum
mass'' model. Podolak (\cite{Podolak03}) used a steady-state model of
an accreting disk to study how the location of the snowline depends on
accretion rate, ice grain size, and contamination of ice by other
materials. He defines the snowline as the location where the rate of
ice grain evaporation equals the rate of ice condensation from
water vapor. This is a more accurate, but still an indirect way of
calculating the location of the snowline. Podolak finds that in the
midplane, where most grains are to be found, the snowline is nearly
independent of grain size and composition, but dependent on
accretion rate.

These efforts do not address evolutionary character of protoplanetary
disk, and, in particular, the decoupling of solids from the gas.
However, the work by Stepinski \& Valageas (\cite{SV97}), recently
confirmed by more computationally sophisticated calculations by
Weidenschilling (\cite{Weidenschilling03}) indicates that the
redistribution of solids is a very important process in the evolution
of the disk. Due to coagulation and sedimentation the solid particles
grow on a time scale that is short in comparison to the lifetime of
the disk. As they grow, they start to drift inward, toward the star,
and as they grow to $\sim 1$ km sizes they settle into a fixed
Keplerian orbit. This ability of the solids to stop migrating inwards
allows for a seemingly unlikely situation in which $R_{sl}$ does not
coincide with $R_{evap}$. The snowline has to be calculated directly
from its definition by keeping track of an evolving distribution of
water-ice.  Calculating the snowline indirectly from gas properties
would result in wrongly placing the snowline too close to the star.

In the present communication we demonstrate the difference between
$R_{sl}$ and $R_{evap}$ for a broad sample of disk models analyzed by
Kornet et al.  (\cite{Kornet01}, \cite{Kornet02}). We identify the
radial drift as the major factor responsible for the final location of
the snowline, and we show how redistribution of solids can enhance the
abundance of solid material leading naturally to an emergence of a
giant planet formation zone.  The basic assumptions of our disk model
and the basic numerical methods employed to solve the equations
governing its evolution are briefly introduced in \S 2. The results of
calculations are presented and analyzed in \S 3, while in \S 4 we
discuss our approach and results in relation to ideas and results
reported in the literature.

\section{Method of calculation}

The protoplanetary disk is modeled as a two-component fluid consisting
of gas and solids. The evolution of the gas component is described by
an analytic solution to the viscous diffusion equation, which gives
the surface density of the gas as a function of radius $r$ and time
$t$ (Stepinski \cite{Stepinski}). The viscosity is given by the usual
$\alpha$ model. The temperature of the gas is calculated in the
thin-disk approximation, assuming vertical thermal balance, according
to equations (2) through (6) in Stepinski (\cite{Stepinski}).
 
Initially the solids are in the form of grains, but because in our
model the solids grow all the way to planetesimals our calculations
handle solid objects with a wide range of sizes. We refer to them
collectively as ``dust'', although term ``solid particles'' and ``dust
particles'' are also used. The model of their evolution includes gas
drag effect, sedimentation, coagulation and evaporation.  Only one
component of dust is considered, in this paper corresponding to
water-ice, which has a sublimation temperature $T_{evap}$ = 150 K and
a bulk density $\varrho$ = 1 g cm$^{-3}$. The main assumptions used in
the calculation are (1) at each radius the particles are all assumed
to have the same size (which, of course, varies in time), (2) all
collisions between particles lead to coagulation, (3) in disk regions
with temperature exceeding $T_{evap}$ all water is in the form of
vapour and evolves at the same radial velocity as the gas component,
(4) initially, in disk regions with temperature below $T_{evap}$, the
particles have sizes equal to $a_{min}=10^{-3}$ cm, (5) the systematic
radial velocity of grains is entirely determined by the effects of gas
drag, (6) the evolution of solids does not affect the evolution of the
gaseous disk The evolution of solids is governed by the set of two
equations (see Kornet et al. (\cite{Kornet01}). The first of them is
the standard continuity equation for $\Sigma_s$.  The second equation
can be interpreted as the continuity equation for size-weighted
surface density of solids $\Sigma_a(r)\equiv a(r)\Sigma_s(r)$ with the
source term accounting for the growth of solid particles due to
collisional aggregation. The radius of a solid particle at a distance
$r$ from the star is denoted by $a(r)$.  The set is solved
numerically. Details of our numerical method is given in Kornet et al.
(\cite{Kornet01}).

The initial conditions are parameterized by the quantities $m_0$ (the
mass of the disk in units of solar mass M$_\odot$), and $j_0$ (the
total angular momentum of the disk in units of $10^{52}$ g cm$^2$
s$^{-1}$). Once $m_0$ and $j_0$ are specified, the analytic solution
of Stepinski (1998) gives the gas surface density
$\Sigma_{g,0}(r)=\Sigma_g(r,t=0)$ that serves as an intial condition
for the evolution of the gaseous disk.

At $t=0$ the ratio $\Sigma_s/\Sigma_g$ and the particle radius $a$ are
independent of $r$ (values of 0.01 and $a_{min}$ = 10$^{-3}$cm are
adopted, respectively). Thus, the initial condition for the evolution
of solid disk is $\Sigma_{s,0}(r)=0.01\Sigma_{g,0}(r)$. We have
checked that the results are not sensitive to a particular selection
of $a_{min}$ by recalculating some models with $a_{min}$ =
10$^{-4}$cm.

\begin{figure}
\resizebox{\hsize}{!}{\includegraphics{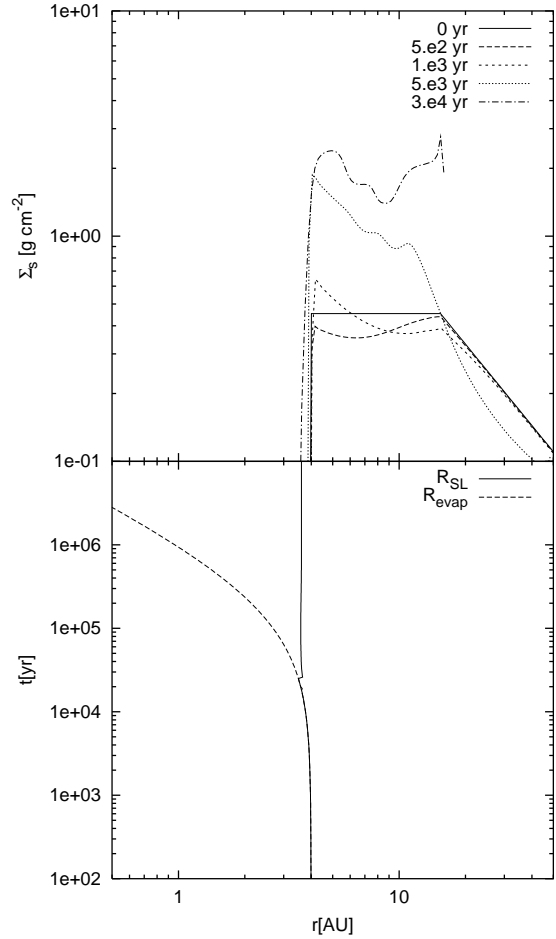}}
\caption{The evolution of solids in an $\alpha$ = 0.1 disk with $m_0$ = 0.02
  and $j_0$ = 1. Top: surface density of solids at the times indicated
  in the frame (solids that have sublimated but not yet accreted onto
  the star are not taken into account).  Bottom: locations of
  sublimation limit (dashed) and snowline (solid) as a function of
  time The snowline is defined as the inner edge of the region where
  ice grains are present.}
\label{fig:disk_example}
\end{figure}
%

\section{Results}
Our results are divided into three grids of disk models. Each grid
groups models with the same constant value of the viscosity parameter
$\alpha$ and contains complete evolutionary information for disks
starting from 99 different initial conditions specified by 0.02 $\le$
$m_0$ $\le$ 0.2 M$_\odot$ and 1$\le$ $j_0$ $\le$ 25. The grids are
characterized by $\alpha=10^{- 3}$, 10$^{-2}$ and 10$^{-1}$,
respectively. The models are evolved until either (1) the outer edge
of the dust disk falls within 0.1 AU, in which case all dust is
assumed to have accreted onto the star, or (2) the total elapsed time
is $10^7$ yr.  In the second case, it usually occurs that the surface
density distribution of solids, $\Sigma_s(r)$, converges to a
stationary configuration well before $10^7$ yr.

\begin{figure*}
\centering
\includegraphics[width=17cm]{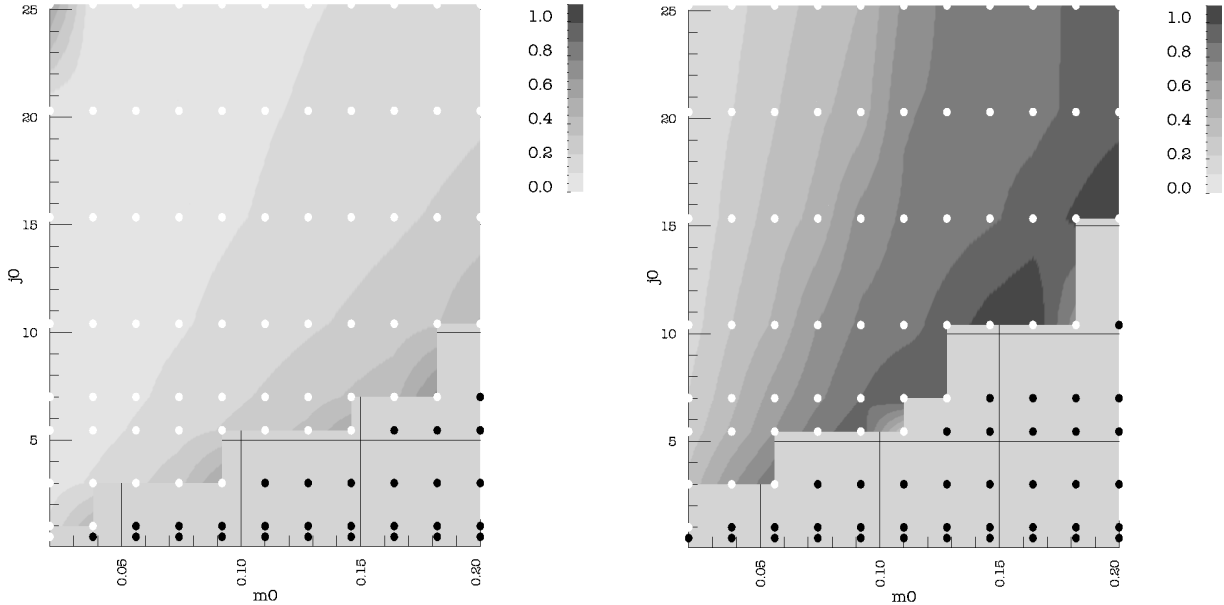}
\caption{The decimal logarithm of surface density enhancement $\xi_s$, defined by equation 
  (\ref{eq:density_enhancement}) for models with $\alpha$ = 0.001
  (left) and $\alpha$ = 0.1 (right). The values of $\log \xi_s$ are
  plotted on the plane of $m_0$ and $j_0$ (the initial values of mass
  and angular momentum of the disk). Within the light-gray region at
  the lower right corner of each frame the solid component has
  completely accreted onto the star. The step-like shape of the
  boundary of that region results from insufficient resolution of the
  grid (the locations of our models are indicated by dots).}
\label{fig:density_enhancement}
\end{figure*}
%

An example of such convergence is illustrated in
Fig.~\ref{fig:disk_example}, where, for a disk with $m_0$ = 0.02,
$j_0$ = 1 and $\alpha$ = 0.1, the surface density of solids is shown
at several selected times as a function of the distance $r$ from the
star. Initially the outer edge of the disk $R_{out}$ and the
sublimation limit $R_{evap}$ are located, respectively, at $\sim$50 AU
and $\sim$4 AU. The surface density of solids drops from about 0.5 g
cm$^{-2}$ at $R_{evap}$ to about 0.1 g cm$^{-2}$ at $R_{out}$.
Interior to $R_{evap}$ the solids remain in the form of vapour, and as
such they are tightly coupled to the gas; however at $r$ $>$
$R_{evap}$ the solid particles begin to drift with respect to the gas.
While drifting, they grow at a rate dependent on $r$ (more slowly at
larger $r$ because of lower densities and collision rates).  By $t$ =
500 yr enough solid material from the outer disk has drifted across
the sublimation limit to cause a noticeable depletion of solids
between $r$ = 4 AU and $r$ = 10 AU. At $t$ = 1000 yr the particles
arriving from the outer disk to the sublimation limit are large enough
to fall onto the descending branch of the drift velocity curve (shown,
for example, in Fig.~1 of Weidenschilling (\cite{Weidenschilling97})),
and as a result a maximum of $\Sigma_s$ develops immediately exterior
to $R_{evap}$. Due to large viscosity, the gaseous component of the
disk evolves so quickly and cools so rapidly that at $\sim5\times10^3$
yr $R_{evap}$ begins to decrease, and the inner edge of the dust disk
moves closer to the star. At about the same time the solid particles
that have arrived at the sublimation limit are so large that they
practically stop drifting. As their motion is essentially Keplerian,
one may say that they form a ``Keplerian barrier'' blocking the flow of
solids across $R_{evap}$. Beyond $\sim$ 15 AU $\Sigma_s$ is now
markedly lower than in the initial model, as the solids originally
residing there have moved closer to $R_{evap}$. An outline of the
final dust disk is already visible in a region between $r$ = 3.5 AU
and $r$ = 15 AU. The evolution is completed at $t$ = $\sim3\times10^4$
yr, when all solids that have not evaporated are collected in that
region. $\Sigma_s$ is now almost uniform, reaching a value $\sim4$
times larger than at the same locations in the initial model. The
typical particle radius $a$ is 300 m. At $t$ = $\sim9\times10^4$ yr
$a=1$ km - i.e. the solids have evolved into planetesimals. At still
later times the planetesimals would tend to grow even further, but the
model is no longer valid because it does not include gravitational
effects. The snowline (i.e. the inner edge of the planetesimals swarm)
stabilizes at $R_{sl}$ = 3.5 AU, while the sublimation limit continues
to shrink, and at $t\sim3\times10^6$ yr it is equal to $\sim$0.5 AU.

Note that the movement of the sublimation limit is always slower than
the velocity of the gas, so the vapor does not leak to the region
beyond $R_{evap}$. In addition, unlike in the model proposed by
Stevenson \& Lunine (\cite{Stevenson}), in our model there is no
significant outward diffusion of water vapor through the
sublimation/condensation radius.  This is because the vapor is
embedded in a dynamically dominant gas and diffuses together with it.
In viscous accretion disk the net result of diffusion of adjunct rings
of gas is to produce an inward mass flow at times and radii relevant
to the sublimation radius. Thus, there is a negligible leak of vapor
to the region $r>R_{evap}$. Overall, in our model, the disk develops a
zone where $T<T_{evap}$, but ice does not exist, as it has been
stopped at larger radii by particle growth and an associated decay of
particle radial drift.

In the above example the final planetesimals swarm forms a
well-defined ring $\sim$3.5 AU $\le$ $r$ $\le$ $\sim$15 AU,with
$\Sigma_s$ in excess of $2$ g cm$^{-2}$. Whether such a swarm leads to
the formation of giant planets cores is unclear as $\Sigma_s$ is lower
than required (see Lissauer \cite{Lissauer}), but the surface density
is relatively constant throughout the entire zone in contrast to the
steep decline ($\Sigma_s \sim r^{-3/2}$) in the ``minimum mass'' model.
Similar structures with different sizes and locations are observed in
all models in which the solids are present at the end of the
evolution. In all models the ratio of final-to-initial $\Sigma_s$ is
greater than 1. The latter property of the final planetesimal swarms
is illustrated in Fig.~\ref{fig:density_enhancement}. To obtain
Fig.~\ref{fig:density_enhancement}, the maximum value of the
enhancement ratio
\begin{equation}
  \xi_s = \frac{\Sigma_s^{max}(t=10^7)}{\Sigma_s(R^{max},t=0)}
  \label{eq:density_enhancement}
\end{equation}
was calculated for each model, and the distribution of $\xi_s$ was
plotted on the ($m_0$, $j_0$) plane. In the above formula
$\Sigma_s^{max}$ is the maximum density found in a given disk, and
$R^{max}$ is the location at which that maximum was found. The
distributions are shown for $\alpha = 10^{-3}$ and $\alpha = 10^{-1}$
(the plot for $\alpha = 10^{-2}$ is omitted because it does not
qualitatively differ from the two cases displayed in
Fig.~\ref{fig:density_enhancement}). In each frame the gray area
indicates the region occupied by disks in which all solids are
accreted onto the star.  The step-like shape of its boundary results
from the poor resolution of the grid (the locations of our models are
indicated by dots). The enhancement ratio $\xi_s$ increases as $j_0$
decreases and $m_0$ increases, i.e. as the initial models become more
compact and dense. This is because in denser disks the solid particles
grow to ``Keplerian'' sizes more quickly, and once the Keplerian barrier
is formed at $R_{evap}$ the solids that still remain at $r>R_{evap}$
are saved from sublimation. At later times the distribution of solids
can only evolve toward a more compact configuration, i.e. $\xi_s$ can
only grow. The earlier the Keplerian barrier forms, the more solids
are locked in the outer disk, the further proceeds the
``compaction'', and the larger may be the final value of $\xi_s$.

Note that for every $m_0$ a critical value $j_{0,c}$ can be chosen
such that disks with $j_0<j_{0,c}$ are entirely void of solids.
The transit from disks with maximum $\xi_s$ to disks in which all
solids accrete onto the star is very abrupt. This is because the
Keplerian barrier is not formed at all when the sublimation limit
falls initially so close to $R_{out}$ that there is no time for the
particles to grow to Keplerian sizes before they arrive at $R_{evap}$.

One can also see that in more viscous disks $\xi_s$ is larger. This is
because the particles in hotter disks gain larger drift velocities.
From eq. (6) in Kornet et al. (\cite{Kornet01}) we can estimate the
maximum inward drift particle velocity
\[
\begin{split}
V_{dr}^{max}& \approx \frac{1}{\rho}\left(\frac{\partial P}{\partial
  r}\right)\frac{r}{V_k} \\
 & \approx \frac{1}{\rho} \left(\frac{d P}{d
  \rho}\right) \left(\frac{\partial \rho}{\partial
  r}\right) \frac{r}{V_k} \\
 & \approx \left(\frac{\partial \ln \rho}{\partial \ln r}\right) \frac{C_s^2}{V_k}
\end{split} 
\]
where $\rho$ is gas density , $P$ -- gas pressure, and $C_s$ -- sound
velocity. As a result, the outer radius of the planetesimal swarm in
hotter disks is smaller, and $\xi_s$ may reach larger values through
more ``compaction''.

\begin{figure}
\resizebox{\hsize}{!}{\includegraphics{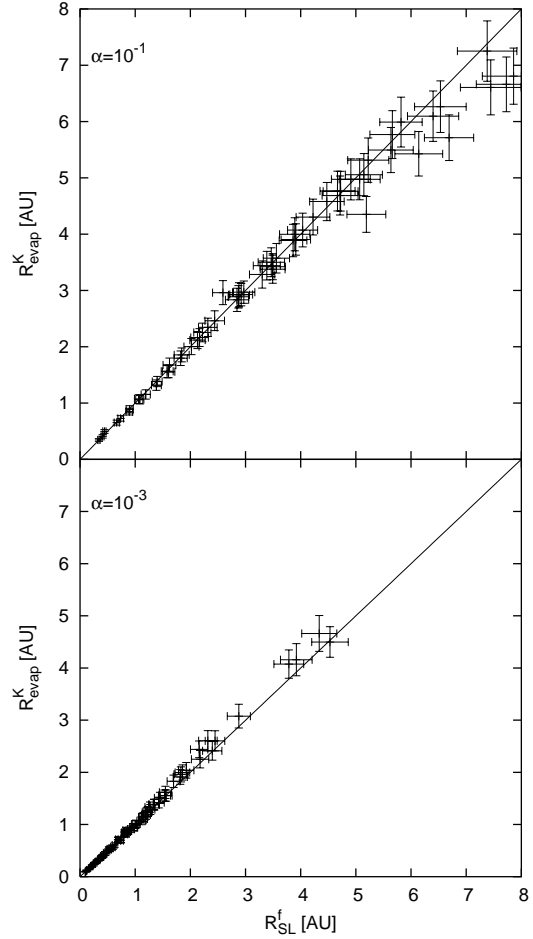}}
\caption{Correlation between the final location of the snowline $R_{sl}^f$ and 
  the distance $R_{evap}^{K}$ at which ``Keplerian particles'' appear at
  the sublimation limit. Points with error bars indicate the location
  of our disk models on the ($R_{sl}$,$R_{evap}^{K}$) plane. The
  length of the error bars is equal to the local grid spacing in $r$
  In the lower frame $R_{sl}^f$ is smaller because disks with smaller
  $\alpha$ are cooler.}
\label{fig:r_in-r_evap}
\end{figure}

Our results indicate that the most important event in the evolution of
the disk occurs when the first ``Keplerian'' particles appear at
$R_{evap}$. Let $R_{evap}^{K}$ be the value of $R_{evap}$ at the
moment the particles arriving at the sublimation limit are so large
that their drift time
\begin{equation}
t_{dr} = \frac{R_{evap}}{v_d\left(a(R_{evap})\right)}
\label{eq:drift_time}
\end{equation}
is equal to 10$^6$ yr. In the above formula $v_d(a)$ is the drift
velocity of particles with a radius $a$, and $a(r)$ is the radius of
particles at a distance $r$ from the star (recall that in our model at
each $r$ the particles are all assumed to have the same size).
Fig.~\ref{fig:r_in-r_evap} shows how well $R_{evap}^{K}$ correlates
with the inner radius of the final planetesimal swarm, i.e. with the
final location of the snowline, $R_{sl}^f$. To explain the correlation
note that as the disk cools, $R_{evap}$ decreases at the rate $\dot
R_{evap}$.

As long as particles at $R_{evap}$ have velocities
$v_d(a(R_{evap}))>\dot R_{evap}$ they move inward across $R_{evap}$
and are destroyed maintaining $R_{sl}$ at $R_{evap}$. However, at
later times, particle at $R_{evap}$ becomes larger, decouple from the
gas and settle into Keplerian orbits. This leads to
$v_d(a(R_{evap}=R^{K}_{evap}))<<\dot R_{evap}$, and while $R_{evap}$
continues to decrease with time, the solids lack inward velocity to
follow it. The finite snowline $R_{sl}^f$ forms approximately at the
location where the inner movement of $R_{sl}$ stalls, that is at
$R_{evap}^{K}$. In disk's outer zone ($r>R_{sl}^f$) the inward motion
of solids continues for some additional time leading to further
``compaction'' of solids.

%

Because many exoplanets have been found relatively close to their
parent stars (see \S 1), it is natural to identify models leading to
the formation of giant planets as close to the star as possible.
Note, however, that these are not necessarly the models with the
smallest value of $R_{sl}^f$. For given values of $\alpha$ and $m_0$,
the models with maximum value of $j_0$ are extended and cool and would
evolve to have smallest possible values of $R_{sl}^f$. However, due to
their low densities, these are not the disks that are expected to
produce planetary cores, thus their snowlines, although located close
to the star, are not of great interest. To identify disk models
forming giant planets at the closest possible distances to the star
would require integration of our models with the models of giant
planet formation, a task that is beyond the scope of the present
paper. Kornet et al.  (\cite{Kornet02}) have probed this issue working
with models similar to those investigated here, but with
high-temperature silicates instead of water-ice being the sole species
of solids. They have found that, for given values of $\alpha$ and
$m_0>m_0^{crit}$, the disks with minimum (but still swarm-producing)
values of $j_0$ are the best candidates to form a close-to-the-star
giant planet at $r \simeq 2$ AU.  With water-ice as the sole species
of solids the results are qualitatively the same, except the inner
edge for giant planet formations is at $r \simeq 4.4$~AU and $\simeq
8.3$ AU for a disk characterized by $\alpha=10^{-3}$ and $10^{-1}$
respectively.

%

\section{Discussion }

In this paper we have concentrated on the issue of the snowline in a
protoplanetary disk in which solids, including water-ice, transform to
ever larger particles through hierarchical coagulation. In such a disk
the solids, initially in the form of dust grains, grow all the way to
planetesimals acquiring a significant inward radial velocities in the
process. However, these velocities are lost when the size of solid
particles approaches planetesimal size, preventing their penetration
into the inner disk regardless of its temperature. This evolution of
solids occurs on a time scale that is short in comparison with the
disk's lifetime. As a result solids decouple from the gas and the
location of the snowline cannot be calculated from properties of the
gaseous disk, such as the gas temperature corresponding to ice
evaporation. Instead, the snowline must be calculated from its direct
definition as the minimum radius at which ice exists.

Using an example of a disk evolving from a set of particular initial
conditions we have demonstrated that although the snowline ($R_{sl}$)
coincides with the location of an ice sublimation/condensation
temperature ($R_{evap}$) during early stages of disk evolution, at
later times $R_{sl}$ remains fixed at $R_{sl}^f$ whereas $R_{evap}$
decreases as the gaseous disk cools. Thus, at these later times,
there exists a zone, $R_{evap}<r<R_{sl}^f$, free of ice despite having
gas temperatures $T<T_{evap}$.

This counter-intuitive result can be understood as follows. In an
accretion disk with the overall inward mass flux, a radial zone is
constantly replenished by material from beyond its outer radius. At
the beginning of the disk's evolution the ``no-ice'' zone described in the
previous paragraph was too hot to support ice. When the temperature in
this zone dropped below the condensation level, the zone no longer
consisted of its original vapor-rich material, instead it consists of
material carried from an outer disk that has been depleted of ice by
decoupling of solids from the gas.

Because we are considering an evolving disk, the evaporation radius is
a function of time, $R_{evap}(t)$, and calculating a unique snowline
using the traditional method of solving equation $T(r,t)=T_{evap}$ is not
possible.  The so defined ``snowline'' would be a function of time and could
be located at arbitrarily small radii. However, $R_{sl}$ calculated
directly from ice presence converges to a single value $R_{sl}^f$ and
constitutes a reasonable estimate for an inner edge of giant planet
formation zone. The location of $R_{sl}^f$ depends on the disk's initial
conditions and the value of $\alpha$. We have demonstrated that the
value of $R_{sl}^f$ is set by the value of $R_{evap}^K$, an
evaporation radius at the moment when solids there are so large that
they settle into Keplerian orbits. Thus, the snowline is determined by
a complicated interplay between coagulation, sedimentation and gas
properties.

We have shown that in the potential giant planet formation zone,
$r>R_{sl}^f$, there is a significant enhancement of density of solids
above its initial value.  This enrichment is due to drift of solids
from an outer disk. All solids are ``compacted'' into a ring between
$R_{sl}^f$ and some outer radius that, however, is significantly
smaller than the outer radius of the gaseous disk. Thus, in our model,
the relatively high surface density of ice can be achieved naturally,
as a result of solids decoupling from the gas.

The above conclusions are based on a simplified description of the
processes governing the evolution of protoplanetary disks, and one may
wonder how robust they are. Below we critically assess several
assumptions on which our model is based.

The most radical is the assumption about the size distribution of
solid particles, which at each distance from the star are assumed all
to have the same diameter. This assumption is responsible for
formation of a ``totally impenetrable'' barrier at $R_{sl}^f$. However,
detailed work by Morfill (\cite{Morfill}) and Weidenschilling
(\cite{Weidenschilling97}) showed that the size distribution of solids
in a protoplanetry disk quickly converges to a stage in which most of
the mass is concentrated in a narrow range of sizes approching the
maximum size. Thus, our approximation can be regarded as a reasonable
idealization. The emergence of an impenetrable barrier in our models
is corroborated by the behavior of solids in the two-dimensional disk
model by Weidenschilling (\cite{Weidenschilling03}) where the incoming
mass tends to pile up at a distance where large bodies form, producing
a sharp transition in both surface density and mean size. We expect
that, in reality, the barrier at $R_{sl}^f$ would leak some ice
particles inward, but they would have a negligible mass.

Our second assumption concerns the 100\% efficiency of coagulation.
In reality, one hardly expects that collisions between solid ice
particles always lead to sticking without fragmentation.
Interestingly, our calculations suggest that if solids in the
protoplanetary disk indeed transform themselves into larger sizes by
hierarchical coagulation, and the efficiency of coagulation is low,
the disk would be depleted of most of its solids, diminishing the
opportunity for planet formation.

To illustrate how the evolution of solids depends on the efficiency of
coagulation we have considered again the particular model described in
\S 3, but with an additional parameter $0\le\epsilon\le1$ regulating
the efficiency of coagulation. To vary this efficiency, we multiply by
$\epsilon$ the source function for particle growth by coagulation,
$f$, (see Kornet et. al., \cite{Kornet01}). The model shown in
Fig.~\ref{fig:disk_example} was obtained for $\epsilon=1$. Decreasing
$\epsilon$ to 0.5 and 0.25 caused the final outer radius of the icy
planetesimals swarm to shrink from 15 AU to 7.5 AU and 2.75 AU,
respectively. The final inner radius moved from 2.65 AU to 2.50 AU and
2.33 AU, respectively. For $\epsilon=0.2$ all solids were accreted
onto the star. This demonstrates how low coagulation efficiency
inhibits development of giant planets formation zone, at least within
a paradigm of hierarchical coagulation.

\begin{figure}[h]
\resizebox{\hsize}{!}{\includegraphics{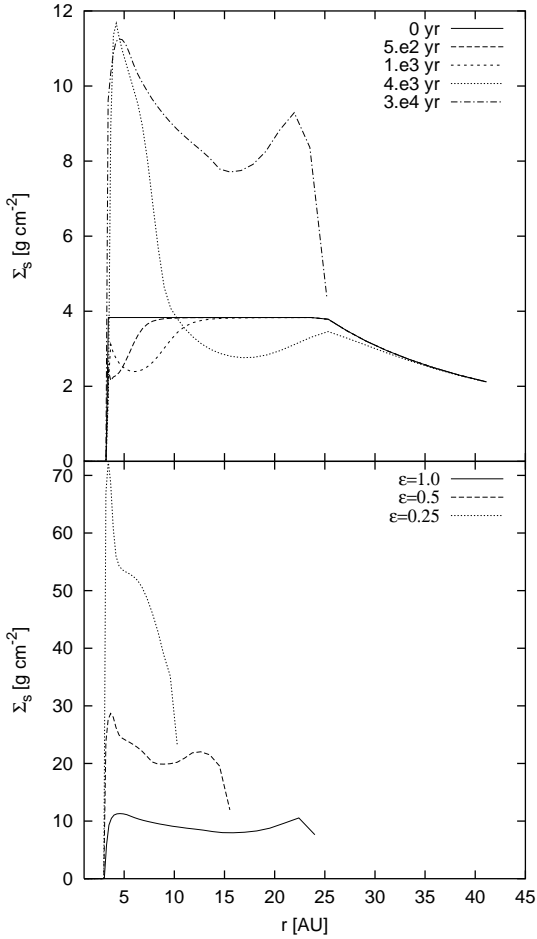}}
\caption{The evolution of solids in an $\alpha$ = 0.001 disk with $m_0$ = 0.2
and $j_0$ = 9. Top: surface density of solids at the times indicated in the 
frame. Bottom: final distributions of solids obtained for the indicated 
vaules of the coagulation efficiency parameter $\epsilon$.}
\label{fig:solar_system}
\end{figure}
%
 
The inferred location of icy planetesimals zone in the early Solar
System is between 3 AU and $\sim$50 AU. The estimation of an inner
edge (i.e. the snowline) is based on dynamical considerations; a
closer snowline could produce Jupiter at close location, which would
prevent formation of terrestrial planets. The estimate of the outer
radius is based on an observation that the Kuiper Belt seems to be
truncated at $\sim50$ AU (Weidenschilling \cite{Weidenschilling03}).
Of all of our models, the model with $\alpha$ = 10$^{-3}$, $m_0$ = 0.2
M$_\odot$, $j_0$ = 9 and $\epsilon=1$ produces the icy planetesimal
swarm closest to the one inferred for the Solar System. The evolution
of solids surface density in such a model is shown in
Fig.~\ref{fig:solar_system}, the final swarm extends from 3 AU to
$\sim25$ AU with surface density ranging from $\sim8$ to $\sim12$ g
cm$^{-2}$.  Note that the solid surface density at 5~AU is above
10~g~cm$^{-2}$, which was the standard value used by Pollack et al.
(\cite{Pollack96}) to form Jupiter in less than 10 Myr.

Our model provides a reasonable scenario for the formation of icy
planetesimal swarms in the Solar System. Note, however, that a good
fit requires $\epsilon\approx 1$. For $\epsilon=0.5$ and
$\epsilon=0.25$ the outer radius of the swarm decreases to 15 AU and
10 AU, respectively, not far enough to account for the Solar System.
Indeed, to account for the entire Kuiper belt, models with $\epsilon$
increasing with the distance from the star, up to values greater than
unity in the outer disk are needed. Note that given the numerous
approximations on which the form of $f$ is based, $\epsilon>1$ is not
as incongruous as it appears.  For example, in low-viscosity disks
($\alpha<\sim0.01$) the relative particle velocities induced by
turbulence become smaller than the differential drift velocities
(Weidenschilling, \cite{Weidenschilling97}). The latter are not
included in $f$, and, as a result, in such a regime our approach
underestimates the particle growth rate which could be countered by
increasing the value of $\epsilon$. Also, in our models the
coagulation rate is underestimated by the absence of other species of
solids. Their presence would increase the solid density and thus
coagulation rate.  This also can be qualitatively countered by
increasing the value of $\epsilon$.

These factors notwithstanding, a high coagulation efficiency is needed
in our model to produce an icy planetesimal swarm leading to a
Solar-like planetary system. However, the high efficiency of
coagulation is questionable.  Indeed, a so-called ``meter-sized
particle barrier'', wherein particles growing to meter-size rubble
achieve high relative velocities and their collisions lead to
fragmentation, is discussed in the literature (see, for example
Weidenschilling \& Cuzzi \cite{Weidenschilling93}). Thus, the demand
for a high coagulation efficiency in our model, to produce planets,
can be view as an argument against a hierarchical coagulation as the
sole agent of solid aggregation.  Perhaps some form of a collective
process to transform solids from dust directly into planetesimals,
thus bypassing the interim stages and avoiding large drift velocities
that lead to the loss of solids, has to take place for the disk to
create planets.  Such a mechanism, in the form of gravitational
instability of the dust layer, was originally proposed by Goldreich \&
Ward (\cite{Goldreich}), but rejected when it became clear that the
presence of turbulence would prevent dust from settling into a thin
enough layer to become gravitationally unstable.  Later, a variant of
the gravitational instability mechanism was proposed wherein the dust
density was enhanced in gaseous vortices to produce clumps dense
enough to be collapsed into planetesimals by gravitational instability
(Tanga et al. \cite{Tanga}). Barring the existence of some collective
process, the paradigm of hierarchical coagulation requires a high
efficiency of coagulation for the formation of giant planets.  This
requirement is independent of the details of our model and is rooted
in the high drift velocities of intermediate-size particles.

Finally, we stress that the models discussed in the present
communication are not meant to be quantitative. Instead, they should be viewed
as an illustration of the basic processes involved, which, we believe, is
qualitatively correct despite its simplicity.  Future improvements of
the model should allow for a range of particle sizes at each position
in the disk and for interactions between different types of particles;
a more detailed description of the gas component is also a desirable
option.

\begin{acknowledgements}
This work was supported in part by the Polish Committee for Scientific 
Research through the grant No. 2P03D01419. KK and MR also benefited from 
the European Commission RTN grant No. HPRN-CT-2002-00308. This is
Lunar and Planetary Institute Contribution No. 1188.

\end{acknowledgements}

\end{document}